\documentstyle[aps,prl,epsf]{revtex}

\begin{document}

\title{On Casimir Forces for Media with Arbitrary Dielectric 
Properties} 
\author{ W. L. Moch\'an}
\address{ Centro de Ciencias F\'{\i}sicas, Universidad Nacional
Aut\'onoma de M\'exico, 
Av. Universidad S/N\\Cuernavaca, Morelos  62210, M\'exico.}
\author{C. Villarreal and R.Esquivel-Sirvent}
\address{Instituto de F\'{\i}sica, Universidad Nacional Aut\'onoma de
M\'exico, Ciudad Universitaria \\ M\'exico , DF 04510, M\'exico.}

\maketitle

\begin{abstract}
We derive an expression for the Casimir force between 
slabs with arbitrary dielectric properties characterized by their 
reflection coefficients. The formalism presented 
here is applicable to media with 
 a local or a
non-local dielectric response, an
infinite or a finite width, inhomogeneous
dissipative, etc. Our results reduce to the Lifshitz formula for the 
force between semi-infinite dielectric slabs by replacing the 
reflection coefficients by the Fresnel amplitudes.  

\end{abstract}


\section{Introduction}

Even though Casimir predicted in 1948 an attractive force between
perfectly conducting plates placed in quantum vacuum,\cite{casimir}
it is only in recent years that experimental studies of Casimir forces
have reached the necessary accuracy to test in detail theoretical
predictions. The first  measurements were done by Derjaguin $et al.$ 
\cite{derja} in 1951 using dielectric materials. In the following 
decades, a number of experiments to measure Casimir interactions between 
dielectric or conducting materials were performed, however involving 
large relative errors in the 
measured forces \cite{milonni}.    
It was until
1997 that Lamoreaux performed measurements with a precision of the 
order of 5 \%  
\cite{lamoreaux} by using an electromechanical system based on a torsion 
balance. 
Other experiments were made taking advantage of the sensitivity of 
atomic force microscopes achieving precisions close to 1\% \cite{mohideen1,mohideen2}.  
Additional measurements have been made by Chan et 
al. \cite{chan} using a micro torsional balance. This experiment is 
representative of the effects that  Casimir forces have in 
micromechanical systems as was theoretically shown by Serry and 
Maclay \cite{serry}. Applications to nanostructures have also been 
considered \cite{nos1,nos2,nos3}. 
This has boosted
investigations in which the detailed properties of the materials such
as absorptivity, rugosity, or finite temperature effects are taken
into account in the theoretical calculations of the Casimir
forces \cite{bordag}.

The standard approach to study vacuum forces between imperfect
conductors is the macroscopic theory proposed by Lifshitz in 1956 for
semi-infinite dielectric materials.\cite{lifshitz} In this theory,
the dissipative effects associated to the radiation reaction of the
elementary atomic dipoles composing the dielectric is balanced by the
fluctuating vacuum field in accordance with the
fluctuation-dissipation theorem. 

For a configuration of two semi-infinite slabs, 
with dielectric
permitivities $\epsilon_1$ and $\epsilon_2$, separated by a gap of width $L$ and permitivity
$\epsilon_3$, the Lifshitz formula for force per unit area is
\begin{equation}
F(L) = -\frac{\hbar}{2 \pi^2 c^3} \int_1^{\infty} dp p^2 
\int_0^{\infty} d \xi \xi^3  \epsilon_3^{3/2} 
\left[ G_1(\xi,p)^{-1} + G_2(\xi,p)^{-1}\right],
\label{Lif}
\end{equation}
with
\begin{equation}
G_1(\xi,p)=\frac{\epsilon_3 s_1 + \epsilon_1 p}{\epsilon_3 s_1 - \epsilon_1 p} 
\frac{\epsilon_3 s_2 + \epsilon_2 p}{\epsilon_3 s_2 - \epsilon_2 p} 
\times e^{2 \xi p \sqrt{\epsilon_3} L/c} -1,
\end{equation}
and
\begin{equation}
G_2(\xi,p)=\frac{ s_1 +  p}{ s_1 -  p} 
\frac{s_2 +  p}{ s_2 -  p} 
\times e^{2 \xi p \sqrt{\epsilon_3} L/c} -1,
\end{equation}
where $\xi=i\omega$ is an imaginary frequency, $p$ and $s_{1,2}$ are defined 
in terms of the momenta parallel and perpendicular to the slabs $k$, 
and $K_{i}^{2}=k^{2}+ \epsilon_{i}(i \xi) \xi^{2}/c^{2}$,  
respectively: $K_{3}=\sqrt{\epsilon_{3}} \xi p/c$ and 
$K_{1,2}^{2}=\epsilon_{3} \xi^{2} s_{1,2}^{2}/c^{2}$. 
Lifshitz theory has been succesfully
employed in a number of experimental situations, and it yields the
Casimir force for perfect conductors. However, as pointed out by Barash and
Ginzburg,\cite{ginzburg} it is not clear how to generalize the theory
to more 
complex problems, such as nonplanar surfaces, multilayer systems,
anisotropic media, etc. Thus, they proposed an alternative
approach to Lifshitz formula. As a dissipative system
does not posses well defined natural frequencies of oscillation, they
introduced an auxiliary system in which the dielectric permitivity
depended only parametrically on the frequency. This procedure enabled
them to calculate the free energy of the field 
as a sum over  allowed states of the system. In turn, the force per unit area 
was obtained as the derivative of the free energy. 
Kats\cite{kats}
re-elaborated the formalism of Barash and Ginzburg by writing the
dispersion law for surface electromagnetic waves in terms of
the reflection amplitudes $r^s$ and $r^p$ of the media
for $s$ and $p$ polarizations. 
Approximating the reflection coefficients in terms of  the frequency and
wavevector dependent surface impedance
$Z(\omega,Q)$ he was able to obtain approximate non-local corrections to the
Casimir force for good conductors. Kats remarked that
dielectrics materials required alternative formulations, as their
reflection coefficients cannot be
expressed merely in terms of surface impedances. This is not
necessarily correct, as an exact relation between surface impedance
and reflection coefficients may indeed be introduced for arbitrary
systems \cite{halevi}. Jaekel and
Reynaud\cite{reynaud} also rederived Lifshitz formula in terms of
reflection 
coefficients for partially transmitting mirrors.  Their expression
reduces to that obtained by Barash and 
Ginzburg.\cite{ginzburg} However, their derivation is not valid when
dissipation is included.

Other approaches have been used to study the 
vacuum fluctuations in the presence of dielectrics 
\cite{kampen,candelas,mat3}.  More recently, the problem of 
quantization in absorbing media and its applications to Casimir 
forces has been considered by Kupiszewska \cite{dorota1,dorota2} and 
also by Matloob \cite{mat1} in the one-dimensional case. 
Interestingly enough,  when temperature effects are 
neglected the expression for the Casimir force in absorbing and 
non-absorbing materials has the same functional form. This fact 
suggests that it is possible to obtain the Casimir force between two 
dispersive and absorbing slabs without the need of quantizing in an 
absorbing medium as was shown by Reynaud et al. \cite{reynaud} using a scattering 
matrix formalism. 

Within the framework of the above discussion, it seems valuable to
present an alternative, very simple derivation of the Casimir force,
valid for materials
with arbitrary dielectric properties. This is the purpose of the
present paper. 

\section{Formalism.}

Consider two  slabs  $a=1,2$ parallel to the $xy$ plane within free space
and separated by a distance $L$ along the $z$-direction, with
inner boundaries at $z_1=0$ and $z_2=L$. We assume that the slabs are
non-chiral, translational invariant and 
isotropic within the $xy$  
plane, but otherwise they may be arbitrary; they could be identical to each
other or different, they might have a local or a
non-local dielectric response, an
infinite or a finite width, they may be opaque or transparent,
dissipative, inhomogeneous, etc.  We want to describe the
electromagnetic field only within vacuum, so we hide all the details of
the field-matter interaction within the slabs in their reflection
amplitudes $r_a^s$ and $r_a^p$ ($a=1,2$). The reflection coefficients
$r_a^\alpha$ are
determined  by the generalized
surface impedances $Z_a^\alpha$ ($\alpha=s,p$) through
\begin{equation}\label{r}
r_a^\alpha = \frac{Z_a^\alpha -Z_0^\alpha}{Z_a^\alpha + Z_0^\alpha}.
\end{equation}
Here $Z_a^\alpha$ is {\em defined} as the quotient
$E^\alpha_{a\parallel}/H^\alpha_{a\parallel}$ of the components
$E^\alpha_{a\parallel}$ and 
$H^\alpha_{a\parallel}$ of the $\alpha$-polarized 
electric and magnetic 
fields evaluated at the $a$-th 
interface for outgoing boundary conditions beyond $z_a$, taken
along appropriately chosen directions parallel to the surface, 
$Z^s_0 = q/k$ and $Z^p_0=k/q$ are
the surface impedances of 
vacuum, $\vec q_{\pm}=(\vec Q,\pm k)$ are the vacuum 
wavevectors with projection $\vec Q$ parallel to the surface and
components $\pm k$ normal to the surface, $q\equiv\sqrt{q_\pm^2}=\omega/c$,
where $\omega$ is the frequency and $c$ the speed of light. The sign
of $k$ is chosen so that $\vec q_+$ propagates (or decays) as $z$
increases. Upon each reflection, $\vec Q$ and $\omega$ are conserved
while the sign of $\pm k$ is reversed. Notice that Eq. (\ref{r}) {\em is
exact} and that no approximation is involved by the
use of our generalized surface impedances, unlike other works 
\cite{mostepanenko,bezerra} that 
use an inappropiate definition of surface impedance. 
 It should be noticed that for local homogeneous
semiinfinite media,
$Z_a^s=q/k_a$ and $Z_a^p=k_a/(\epsilon_a q)$, where $k_a$ is the
component of the wavevector normal to the surface within medium $a$
with local dielectric response $\epsilon_a(\omega)$, and
Eq. (\ref{r}) yields 
the well known Fresnel amplitudes. However, Eq. (\ref{r}) is much more
general.\cite{halevi}

The density of states within vacuum may be obtained from the
Green's functions of the system. To this end, we study first the case of
$s$-polarized waves choosing $x-z$ as the plane of incidence.
With that choice, ${\vec E}=
(0,E_y,0)$, ${\vec B}= (B_x,0,B_z)$,   ${\vec q_\pm}=
(Q,0,\pm k)$ and the boundary conditions for $\vec E$ become $i q E_y(0^+)=
-Z_1^s \partial_z E_y(0^+)$, and $i q E_y(L^-)= 
Z_2^s \partial_z E_y(L^-)$, with $\partial_z\equiv\partial/\partial
z$.  The electric Green's function is 
\begin{equation}\label{ge}
G^E_{k^2}(z,z') = \frac{E_y^<(z_<) E_y^>(z_>)}{W} ,
\end{equation}
where $z_<$ and $z_>$ are the smaller and larger of $z$ and $z'$
respectively, 
\begin{equation}\label{eym}
E_y^<(z)= e^{-ikz}+ r_1^s e^{ikz}
\end{equation}
and
\begin{equation}\label{eyM}
E_y^>(z)= e^{ik(z-L)}+ r_2^s e^{-ik(z-L)}
\end{equation}
are solutions of the scalar 1D wave equation with wavenumber $k$ obeying
the appropriate boundary conditions at $z=0^+$ and $L^-$ respectively,
and $W$ is their Wronskian. 
Analogously, $\vec B$ obeys
$iq B_x(0^+)= -(Z^s_0)^2 \partial_z B_x(0^+)/Z_1^s$, and 
$iq B_x(L^-)= (Z^s_0)^2 \partial_z B_x(L^-)/Z_2^s$, so that the
magnetic Green's function is obtained by replacing $E_y\to B_x$ and
$r_a^s \to -r_a^s$ in
Eqs. (\ref{ge})-(\ref{eyM}). We do not consider $B_z$ separately, as
it is simply proportional to $E_y$.

For each $\vec Q$, the local density of states per unit $k^2$ is given
by\cite{plunien} 
\begin{equation}\label{rho2}
\rho^s_{k^2}(z) = -\frac{1}{2\pi} \mbox{Im} \left( G^E_{\tilde
k^2}(z,z)+ G^B_{\tilde k^2}(z,z)\right),\quad(\tilde k\equiv k+i0^+) 
\end{equation}
so that by substituting Eqs. (\ref{ge})-(\ref{eyM}) and its magnetic
analogues we obtain
\begin{equation}\label{rhos2}
\rho^s_{k^2}= \frac{1}{2 \pi \tilde k} \mbox{Re} \left[ 
\frac{1+r_1^s r_2^s e^{2i\tilde kL}}{1-r_1^s r_2^s e^{2i\tilde kL}} \right],
\end{equation}
independent of $z$.
The density  $\rho^p_{k^2}$ corresponding to $p$
polarization may be derived similarly, and is simply given by
Eq. (\ref{rhos2}) after replacing all the superscripts $s\to p$.
Finally, the total density of states is  $\rho_{k^2} = \rho^s_{k^2} +
\rho^p_{k^2}$.

A photon in a state characterized by
$\alpha,{\vec Q}$ and $k^2$ has momentum $\pm \hbar k$  and moves
with velocity $ \pm ck/q$ along the $z$ direction, so that its 
contribution to the momentum flux is $\hbar c k^2/q$. Multiplying this by the
photon occupation number, integrating over $k^2$ with the weight
function $\rho_{k^2}^\alpha$ and adding the contributions from  all
values of  $\alpha=s,p$ and 
$\vec Q$, with the usual replacement $\sum_{\vec Q}\ldots\to {\it A}/2\pi \int QdQ\ldots$,
we obtain the momentum flux from the vacuum gap into
slab 2.  There is a similar contribution coming from the semiinfinite
vacuum on the other side of the
slab, obtained by substituting $r_2^\alpha\to0$ above and reversing
the flux direction $z\to-z$. The total force per unit area is 
obtained by subtracting the contributions from either side \cite{pollo}
\begin{equation}\label{F}
F(L)={\cal A}\frac{\hbar c}{2 \pi ^2} \int_0^{\infty} dQ Q \int_{q\ge 0} dk 
\frac{\tilde k^2}{q} \mbox{Re} \left[ 
\frac{r_1^s r_2^s e^{2i\tilde kL}}{1-r_1^s r_2^s e^{2i\tilde kL}} 
+ \frac{r_1^p r_2^p e^{2i\tilde kL}}{1-r_1^p r_2^p e^{2i\tilde kL}} 
\right].
\end{equation}
The integral over $k$ runs from $iQ$ to 0 and then to $\infty$, so
that $q$ remains real and positive.
It is easy to show that for perfect mirrors ($r_a^\alpha=\pm 1$)
Eq. (\ref{F}) yields the expected Casimir 
force. Furthermore,  simple substitution of the Fresnel amplitudes 
\begin{eqnarray} 
    r^{s}_{a}&=&\frac{k-k_{a}}{k+k_{a}}  \\
    r^{p}_{a}&=&\frac{k_{a}-\epsilon_{a}k}{k_{a}+\epsilon_{a}k},
    \end{eqnarray}
and
manipulation of the integration contours in Eq. (\ref{F}) leads to 
the formula of Lifshitz (Eq.(\ref{Lif})).

\section{Conclusions}

We have derived a general expression for Casimir forces between slabs 
with arbitrary dielectric properties characterized by the reflection 
coefficients of the material. This procedure avoids complications 
related to the quantization of the electromagnetic field in dispersive 
media. The expression we obtain for the Casimir force is convenient for 
calculations since the reflection coefficients can be obtained 
straightforwardly in theoretical computations or through experimental 
studies. Our approach is based on the exact definitions of surface 
impedance and yields the Lifshitz formula for semi-infinite slabs 
when the reflection coefficients are replaced by the Fresnel 
amplitudes. This contradicts the results obtained by Mostepanenko 
and Trunov \cite{mostepanenko} and  also by Bezerra {\it et al.} \cite{bezerra} 
that claim that the use 
of surface impedances is only an approximation valid for small 
transverse wave vector $\bf{Q}$. However, this conclusion arises from
considering an approximate expression of the surface impedance.
In our work this limitation is not 
present, since we do take into account that the surface 
impedances for the p and s polarized waves are different. Therefore, the results 
that we obtain are valid for any value of the wave vector $\bf{Q}$.  
The expresion for the Casimir force in Eq. (10) can be used to 
calculate accurately the force between non-homogeneous systems. In fact, current experimental
setups for measuring Casimir forces consist of two metallic surfaces with a 
high reflectivity in the frequency range of interest $\omega \sim c/a$, coated by 
thin ($\sim 8$ nm) Au/Pd layer to avoid Al oxidation. The analysis of 
the experimental data is performed by giving arguments on the 
transparency of the Au/Pd layer, which allows the  
use of the Lifshitz expression for semi-infinite homogeneous media 
with semi-empirical corrections for the presence of that layer.
However, within our formalism, such assumptions are unecessary. In a 
future work, we will present exact results for Casimir forces 
between heterostructured media, generalizing our previous 
one-dimensional analysis \cite{nos1}.

Acknowledgments:
This work was partially supported by NASA Breakthrough Propulsion
Physics Project, and by DGAPA-UNAM Project IN-110999.

\end{document}